\documentclass[12pt]{article}
\usepackage[left=2.54cm,top=2.54cm,right=2.54cm,bottom=2.54cm]{geometry}
\usepackage{amsmath, amssymb}
\usepackage[latin1]{inputenc}
\usepackage{mathrsfs}
\usepackage{cite}
\usepackage{dsfont}

\begin{document}
\date{}
\title{$SU(1,1)$ solution for the Dunkl-Coulomb problem in two dimensions and its coherent states}

\author{M. Salazar-Ram\'{\i}rez$^{a}$\footnote{{\it E-mail address:} escomphysics@gmail.com.mx}, D. Ojeda-Guill\'en$^{a}$ and \\ R. D. Mota$^{b}$} \maketitle

\begin{minipage}{0.9\textwidth}

\small $^{a}$ Escuela Superior de C\'omputo, Instituto Polit\'ecnico Nacional,
Av. Juan de Dios B\'atiz esq. Av. Miguel Oth\'on de Mendiz\'abal, Col. Lindavista,
Del. Gustavo A. Madero, C.P. 07738, Ciudad de M\'exico, Mexico.\\

\small $^{b}$ Escuela Superior de Ingenier{\'i}a Mec\'anica y El\'ectrica, Unidad Culhuac\'an,
Instituto Polit\'ecnico Nacional, Av. Santa Ana No. 1000, Col. San
Francisco Culhuac\'an, Del. Coyoac\'an, C.P. 04430, Ciudad de M\'exico, Mexico.\\

\end{minipage}

\begin{abstract}

We study the radial part of the Dunkl-Coulomb problem in two dimensions and show that this problem possesses the $su(1,1)$ symmetry.
We introduce two different realizations for the $su(1,1)$ Lie algebra and use the theory of irreducible representations  to obtain the energy spectrum and the eigenfunctions. For the first algebra realization, we apply the Schr\"odinger factorization to the radial part of the Dunkl-Coulomb problem to construct the algebra generators.  In the second realization, we introduce three operators, been one of them proportional to the radial Hamiltonian. Finally, we use the $su(1,1)$  Sturmian basis of one of the two algebras to construct the radial coherent states in a closed form.

\end{abstract}

\section{Introduction}

The Dunkl operator was introduced by Yang in the context of the deformed oscillator \cite{YANG}, and Dunkl as part of a program on polynomials in several variables with reflection symmetries related to finite reflection groups (or equivalently finite Coxeter groups) \cite{DUNKL}. The Dunkl operators $D_i$ are combinations of differential and difference operators, associated to a finite reflection group $\mathcal{G}$. This operator allows the construction of a Dunkl Laplacian, which is a combination of the classical Laplacian in $\mathds{R}^n$ with some difference terms, such that the resulting operator is only invariant under $\mathcal{G}$ and not under the whole orthogonal group \cite{DUNKL,XuY}.

The Dunkl operators (involving reflection operators),  are very useful in the study of special functions with root systems and they are closely related to certain representations of degenerate affine Hecke algebras \cite{Chere,Opdam}. The commutative algebra generated by these operators has been used to study integrable models of quantum mechanics, as the Calogero-Sutherland-Moser models \cite{HIK,KAK,LAP}.

The isotropic Dunkl oscillator in two and three dimensions is another important problem in quantum mechanics involving the Dunkl operator. In references  \cite{GEN1,GEN2,GEN3}, the authors showed that they are superintegrable systems and are closely related to the $-1$ orthogonal polynomials of the Bannai-Ito scheme. Also, they obtained the symmetry of the problem (called Schwinger-Dunkl algebra) and the exact solutions of the Schr\"odinger equation in terms of Jacobi, Laguerre and Hermite polynomials. In reference \cite{NOS}, we solved the Dunkl-oscillator problem algebraically by using the $su(1,1)$ Lie algebra and the theory of unitary irreducible representations.

Recently the Dunkl-Coulomb system has been studied in terms of the Dunkl Laplacian \cite{GEN4}. It was shown that this model is both superintegrable and exactly solvable. Moreover, the constants of motion and the symmetry algebra were obtained. The solution for this system was computed in polar coordinates and the symmetry operators were constructed by generalizing the Runge-Lenz vector.

On the other hand, since Dirac \cite{DIRLIB}, Schr\"odinger\cite{SCH1A,SCH1B,SCH1C}, and Infeld and Hull \cite{INF1,INF2} established the fundamental ideas,  the factorization methods have been of great interest for the study of quantum systems. The success of the factorization methods lies on the fact that  if the Schr\"odinger equation is factorable, the energy spectrum and eigenfunctions are obtained in an algebraic way. Moreover, the operators constructed from these methods are related to compact and non-compact Lie algebras.

The coherent states introduced by Schr\"odinger \cite{SCH} as the most classical ones of the harmonic oscillator (those of minimal uncertainty), have been successfully constructed for many physical problems \cite{PERL,ZHANG,GAZL,KLIL}. The works of Barut \cite{AOB} and Perelomov \cite{AMP} generalized the concept of coherent state for any algebra of a symmetry group. Regarding the Perelomov coherent states for the $su(2)$ and $su(1,1)$ Lie algebras several works have been published, as can be seen in references \cite{KWO,CBR,CBR2}.

The aim of the present work is to study the Dunkl-Coulomb problem in algebraic way. In order to obtain the energy spectrum and the eigenfunctions of this problem we use the theory of unitary representations and two algebraic methods: the Schr\"odinger factorization and the tilting transformation.

This work is organized as follows. In Section $2$ we obtain the Hamiltonian for the radial and angular part of the Dunkl-Coulomb problem in two dimensions in polar coordinates. In Section $3$, we apply the Schr\"odinger factorization method for the radial Hamiltonian to obtain the three generators of the $su(1,1)$ Lie algebra. The energy spectrum is computed by using the theory of irreducible representations and the eigenfunctions are obtained analytically. In Section $4$, the energy spectrum and the eigenfunctions are obtained by using the tilting transformation and a realization of the $su(1, 1)$ Lie algebra which is energy-independent. In Section $5$, we compute the $SU(1,1)$ Perelomov coherent states for the radial part of the Dunkl-Coulomb problem in the plane. Finally, we give some concluding remarks.

\section{The Dunkl-Coulomb problem in two dimensions in polar coordinates}

The Hamiltonian for the Dunkl-Coulomb problem is written as
\begin{equation}\label{HDC}
H=-\frac{1}{2}\nabla_D^2+\frac{\alpha}{r},
\end{equation}
where $r^2=x_1^2+x_2^2$ and $\nabla_D^2$ is the two-dimensional Dunkl-Laplace operator given by
\begin{equation}
\nabla_D^2=D_1^2+D_2^2.
\end{equation}
The Dunkl derivative $D_i$ is defined as
\begin{equation}
D_i=\partial_{x_i}+\frac{\mu_i}{x_i}(1-R_i), \hspace{1.0cm} i=1,2.
\end{equation}
In this expression $R_i$ is the reflection operator
\begin{equation}
R_if(x_i)=f(-x_i),
\end{equation}
with respect to the $x_i=0$ axis and $\mu_i>0$ are real parameters.

The Schr\"odinger equation $H\Psi=\mathcal{E}\Psi$ for the Dunkl-Coulomb problem is  exactly solved using separation of variables in polar coordinates $(x_1=r\cos\phi,x_2=r\sin\phi)$. For this coordinates, the reflection operators $R_1$ and $R_2$ have the actions
\begin{equation}
R_1f(r,\phi)=f(r,\pi-\phi), \hspace{1.0cm} R_2f(r,\phi)=f(r,-\phi).
\end{equation}
Thus, the Hamiltonian in polar coordinates of equation (\ref{HDC}) is written as
\begin{equation}\label{Hamr}
H=A_r+\frac{1}{r^2}B_\phi
\end{equation}
where $A_r$ and $B_\phi$ are given by \cite{GEN4}
\begin{align}\label{Ar}
A_r &=-\frac{1}{2}\partial_r^2-\frac{1}{2r}(1+2\mu_1+\mu_2)\partial_r+\frac{\alpha}{r},\\\label{Bp}
B_\phi &=-\frac{1}{2}\partial_\phi^2-(\mu_1\tan\phi+\mu_2\cot\phi)\partial_\phi+\frac{\mu_1(1-R_1)}{2\cos^2\phi}+\frac{\mu_2(1-R_2)}{2\sin^2\phi}.
\end{align}
Therefore, by using the equations (\ref{Hamr}),(\ref{Ar}) and (\ref{Bp}) the Schr\"odinger equation of the radial and angular part are
\begin{align}\label{Erad1}
&\left(A_r-\mathcal{E}+\frac{s^2}{2r^2}\right)R(r)=0,\\
&\left(B_\phi-\frac{s^2}{2r^2}\right)\Phi(\phi)=0,
\end{align}
where it has been proposed $\Psi=R(r)\Phi(\phi)$ and $s^2$ is the separation constant given by
\begin{equation}\label{scuad}
s^2=4m(m+\mu_1+\mu_2).
\end{equation}

The solution for the angular part is given in terms of the quantum numbers ($e_1,e_2$) corresponding to the eigenvalues($1-2e_1,1-2e_2$) of the reflection operators ($R_1,R_2$) ($e_i\in\left\{0,1\right\}$). Thus, the solution for the equation (\ref{Bp}) is \cite{GEN4}
\begin{equation}
\Phi_n^{(e_1,e_2)}(\phi)=\eta_n^{(e_1,e_2)}\cos^{e_1}\phi\sin^{e_2}\phi P_{n-e_1/2-e_2/2}^{\mu_1-1/2+e_1,\mu_2-1/2+e_2}(-\cos{2\phi}),
\end{equation}
where $P_n^{(\alpha,\beta)} (x)$ are the classical Jacobi polynomials. Also, when
\begin{equation*}
  (e_1,e_2)\in\left\lbrace
  \begin{array}{l}
      \left\{(0,0),(1,1)\right\},\hspace{1.0cm}\text{$n$ is a non-negative integer }.\\
      \left\{(1,0),(0,1)\right\},\hspace{1.0cm}\text{$n$ is a positive half-integer }.\\
  \end{array}
  \right.
\end{equation*}
The normalization constant $\eta_n^{(e_1,e_2)}$ is given by
\begin{align}
\eta_n^{(e_1,e_2)}&=\sqrt{\left(\frac{2n+\mu_1+\mu_2}{2}\right)\left(n-\frac{e_1+e_2}{2}\right)}\times\\
&\sqrt{\frac{\Gamma\left(n+\mu_1+\mu_2+\frac{e_1+e_2}{2}\right)}{\Gamma\left(n+\mu_1+\frac{1+e_1+e_2}{2}\right)}\Gamma\left(n+\mu_1+\frac{1+e_2-e_1}{2}\right)}.
\end{align}
From the orthogonality relation of the Jacobi polynomials, it can be deduced
that the radial part of the Dunkl-Coulomb problem satisfies \cite{GEN4}
\begin{equation}
\int_0^{2\pi}\Phi_n^{(e_1,e_2)}(\phi)\Phi_{n'}^{(e'_1,e'_2)}(\phi)|\cos{\phi}|^{2\mu_1}|\sin{\phi}|^{2\mu_2}d\phi=\delta_{n,n'}\delta_{e_1,e'_1}\delta_{e_2,e'_2}.
\end{equation}
In the following two Sections we shall obtain the radial eigenfunctions and the energy spectrum of the Dunkl-Coulomb problem by using two different algebraic methods, the Schr\"odinger factorization and the tilting transformation.

\section{The Schr\"odinger factorization method}

In this Section we shall use the Schr\"odinger factorization method \cite{SCH1A,DANIEL} to obtain the energy spectrum and the eigenfunctions of the Dunkl-Coulomb problem. Thus, if we focus on equations (\ref{Ar}) and (\ref{Erad1}) the radial part for this problem can be rewritten as
\begin{equation}
\left(-r^2\frac{d^2}{dr^2}-r(1+2\mu_1+2\mu_2)\frac{d}{dr}-2\mathcal{E}r^2+2\alpha{r}\right)\mathscr{F}(r)=-s^2\mathscr{F}(r).\label{ecudi2a}
\end{equation}
The Schr\"odinger factorization can be applied to the left-hand side of equation (\ref{ecudi2a}). We propose
\begin{equation}\label{sch}
\left(r\frac{d}{dr}+ar+b\right)\left(-r\frac{d}{dr}+cr+f\right)\mathscr{F}(r)=g\mathscr{F}(r),
\end{equation}
where $a$, $b$, $c$, $f$ and $g$ are constants to be determined. Expanding this
expression and comparing it with equation (\ref{ecudi2a}) we obtain
\begin{align}
a&=c=\pm\sqrt{-2\mathcal{E}},\quad f=\pm\frac{\alpha}{\sqrt{-2\mathcal{E}}}-\frac{1+2\mu_1+2\mu_2}{2},\\ b&=\pm\frac{\alpha}{\sqrt{-2\mathcal{E}}}+\frac{1+2\mu_1+2\mu_2}{2}-1, \\
g&=\left(\frac{\alpha}{\sqrt{-2\mathcal{E}}}\mp\frac{1}{2}\right)^2-(\mu_1+\mu_2)^2-s^2.
\end{align}
Therefore, the differential equation for $\mathscr{F}(r)$ (equation (\ref{ecudi2a})) is factorized as
\begin{align}
\left(\mathscr{A}_\mp\mp1\right)\mathscr{A}_\pm \mathscr{F}(r)
=\left[\left(\frac{\alpha}{\sqrt{-2\mathcal{E}}}\mp\frac{1}{2}\right)^2-\left(\mu_1+\mu_2\right)^2-s^2\right]\mathscr{F}(r),\label{Facdesc1}
\end{align}
where
\begin{equation}
\mathscr{A}_{\pm}=\left(\mp r\frac{d}{dr}+\sqrt{-2\mathcal{E}}r+\frac{\alpha}{\sqrt{-2\mathcal{E}}}\mp\frac{1+2\mu_1+2\mu_2}{2}\right),
\end{equation}
are the Schr\"odinger operators. Thus, the operators $\mathscr{L_{\pm}}$, $\mathscr{L}_0$,  which close the $su(1,1)$ Lie algebra (see Appendix), are given by
\begin{equation}
\mathscr{L}_{\pm}=\left[\mp r\frac{d}{dr}+\sqrt{-2\mathcal{E}}r\mp\frac{1+2\mu_1+2\mu_2}{2}-\mathscr{L}_0\right],
\end{equation}
with the operator  $\mathscr{L}_0$ defined as
\begin{align}\nonumber
\mathscr{L}_0\mathscr{F}(r)& \equiv \frac{1}{\sqrt{-2\mathcal{E}}}\left[-r\frac{d^2}{dr^2}-(1+2\mu_1+2\mu_2)\frac{d}{dr}-2\mathcal{E}r+\frac{s^2}{r}\right]\mathscr{F}(r)\\
&=-\frac{\alpha}{\sqrt{-2\mathcal{E}}}\mathscr{F}(r).\label{B0}
\end{align}
The last equality is due to equation (\ref{ecudi2a}). The quadratic Casimir operator $\mathscr{C}^2$ (equation (\ref{cas}) of Appendix) satisfies the eigenvalue equation
\begin{equation}\label{opcas1}
\mathscr{C}^2\mathscr{F}(r)=\left[(\mu_1+\mu_2)^2+s^2-\frac{1}{4}\right]\mathscr{F}(r)=k(k-1)\mathscr{F}(r).
\end{equation}
By substituting equation (\ref{scuad}) into equation (\ref{opcas1}) we obtain that the group number $k$ (Bargmann index) is
\begin{equation}
k=2m+\mu_1+\mu_2+\frac{1}{2}, \hspace{1.0cm} k=-2m-\left(\mu_1+\mu_2\right)+\frac{1}{2}.\label{k}
\end{equation}
The other group number, $n$,  can be identified with the radial quantum number $n_r$. Thus, the energy spectrum for this problem can be computed from equations (\ref{k0n}) and (\ref{B0}) to obtain
\begin{equation}\label{ener1}
\mathcal{E}=-\frac{\alpha^2}{2(n+2m+\mu_1+\mu_2+\frac{1}{2})^2}.
\end{equation}

The radial function $\mathscr{F}(r)$ can be easily obtained from the general differential equation \cite{LEB}
\begin{equation}
xu''+\left(\sigma+1-2\nu\right)u'+\left[n+\frac{\sigma+1}{2}-\frac{x}{4}+\frac{\nu(\nu-\sigma)}{x}\right]u=0,\label{diflev}
\end{equation}
which has the particular solution
\begin{equation}
u=N_ne^{-x^2/2}x^\nu L_n^{\sigma}(x).
\end{equation}

By making the change of variable $x=2ar$ in the radial equation (\ref{ecudi2a}), and comparing it with the differential equation (\ref{diflev}), we obtain the relationships
\begin{align}\nonumber
\sigma+1-2\nu &=1+2\mu_1+2\mu_2,&n+\frac{\sigma+1}{2}&=-\frac{\alpha}{a},\hspace{0.3cm}\\
\frac{\mathcal{E}}{2a^2}&=-\frac{1}{4}, &\nu(\nu-\sigma)&=-s^2.
\end{align}
Hence, the radial wave functions $\mathscr{F}(r)$ explicitly are
\begin{align}\nonumber
\mathscr{F}(r)&=\sqrt{\frac{(2a)^{2\mu_1+2\mu_2+2}n_r!}{\Gamma(n_r+4m+2\mu_1+2\mu_2+1)(2n_r+4m+2\mu_1+2\mu_2+1)}}\\
&\times e^{-ar}(2ar)L_{n_r}^{4m+2\mu_1+2\mu_2}(2ar).
\end{align}
The normalization coefficient $N_n$ was computed from the orthogonality of the Laguerre polynomials
\begin{equation}
\int_0^{\infty}e^{-x}x^{\alpha}\left[L_{n}^{\alpha}(x)\right]^2dx=\frac{\Gamma(n+\alpha+1)}{n!}.\label{norm}
\end{equation}

Therefore, we have showed that the radial part of the Dunkl-Coulomb problem in polar coordinates possesses the $su(1,1)$ symmetry.
However, unlike other factorization methods, the Schr\"odinger factorization allowed us to construct systematically the $su(1,1)$ generators of the Dunkl-Coulomb problem. These generators and the theory of the $su(1,1)$ Lie algebra were used to obtain the energy spectrum.

\section{The tilting transformation method}

The Dunkl-Coulomb problem can be studied from an alternative $SU(1,1)$ approach called the
tilting transformation. This method is a group theoretical interpretation of scaling from the active viewpoint
and converts a conventional Hamiltonian eigenvalue problem into a $K_0$ eigenvalue problem \cite{ADAMS}.

The tilting transformation method is based on the dilation operator $e^{i\theta K_2}$, where
$\theta$ is a real number. Thus, we introduce the set of operators
\begin{align}\label{A1}
A_0&=\frac{1}{2}\left(-r\frac{d^2}{dr^2}-2\left(1/2+\mu_1+\mu_2\right)\frac{d}{dr}+\frac{s^2}{r}+r\right),\\\label{A2}
A_1&=\frac{1}{2}\left(-r\frac{d^2}{dr^2}-2\left(1/2+\mu_1+\mu_2\right)\frac{d}{dr}+\frac{s^2}{r}-r\right),\\\label{A3}
A_2&=-ir\left(\frac{d}{dr}+\frac{1}{r}\left(1/2+\mu_1+\mu_2\right)\right).
\end{align}
These new operators also close the $su(1,1)$ Lie algebra. They are a generalization of those introduced by Barut
to study general central potentials \cite{AOB2,KTH}. Also, unlike the Schr\"odinger factorization operators,
these operators do not depend on the energy. A direct calculation shows that the Casimir operator for this algebra is
$\mathcal{C}=$$A_0^2-A_1^2-A_2^2$=$s^2-\frac{1}{4}+(\mu_1+\mu_2)^2 $, which results to be equal to the Casimir operator for the algebra
we have introduced in the preceding Section. This implies that the Bargmann index for the above algebra takes the values given
in equation (\ref{k}).

The radial equation for the Dunkl-Coulomb problem $H_r$
\begin{equation}
H_rR(r)=\left[-\frac{1}{2}r\frac{d^2}{dr^2}-\left(1/2+\mu_1+\mu_2\right)\frac{d}{dr}+\frac{s^2}{2r}+\alpha-\mathcal{E}r\right]R(r)=0,
\end{equation}
is written in terms of the $su(1,1)$ Lie algebra generators as
\begin{equation}
H_rR(r)=\left[\frac{1}{2}\left(A_0+A_1\right)-\frac{1}{2}\mathcal{E}\left(A_0-A_1\right)+\alpha\right]R(r)=0.\label{radial}
\end{equation}
This radial Hamiltonian $H_r$ can be diagonalized via the scaling operator by introducing the similarity transformation $\widetilde{R}(r)=e^{-i\theta A_2}R(r)$ and $\widetilde{H}_r=e^{-i\theta A_2}H_re^{i\theta A_2}$. The action of the scaling or tilting transformation onto the operators $A_0$ and $A_1$ can be computed from the Baker-Campbell-Hausdorff formula
\begin{align}
e^{-i\theta A_2}A_0e^{+i\theta A_2}&=A_0\cosh(\theta)+A_1\sinh(\theta),\\
e^{-i\theta A_2}A_1e^{+i\theta A_2}&=A_0\sinh(\theta)+A_1\cosh(\theta).
\end{align}
From these equations, it is easy to show the following property
\begin{equation}
e^{-i\theta A_2}(A_0\pm A_1)e^{i\theta A_2}=e^{\pm\theta}(A_0\pm A_1).
\end{equation}

Thus, from above expression the equation (\ref{radial}) can be written as
\begin{equation}\label{delt1}
\widetilde{H}_r\widetilde{R}(r)=\left[\left(\frac{1}{2}e^{\theta}-\mathcal{E}e^{-\theta}\right)A_0
+\left(\frac{1}{2}e^{\theta}+\mathcal{E}e^{-\theta}\right)A_1-\alpha\right]\widetilde{R}(r)=0.
\end{equation}
If we choose the scaling parameter as $\theta=\ln(-2\mathcal{E})^{1/2}$, the coefficient of $A_1$ vanishes and the tilted
Hamiltonian $\widetilde{H}_r$ is
\begin{equation}\label{ecde}
\widetilde{H}_r\widetilde{R}(r)=\left[(-2\mathcal{E})^{1/2}A_0+\alpha\right]\widetilde{R}(r)=0.
\end{equation}
From the action of the operator $A_0$ on the $su(1,1)$ states (equation (\ref{k0n}) of Appendix) we obtain
\begin{equation}
(-2\mathcal{E})^{1/2}\left(n+k\right)=-\alpha,
\end{equation}
Therefore, by using equation (\ref{k}) for the values of $k$,  we obtain
the energy spectrum of the Dunkl-Coulomb problem
\begin{equation}
\mathcal{E}=-\frac{\alpha^2}{2\left(n+2m+\mu_1+\mu_2+\frac{1}{2}\right)^2}.\label{En}
\end{equation}

The states for the unitary irreducible representations (Sturmian basis) for the Kepler-Coulomb $su(1,1)$ Lie algebra
has been solved by Gerry and Kiefer \cite{gerry} and Gur and Mann \cite{gur}. Thus, we can take advantage of these results to obtain the
Sturmian basis for the Dunkl-Coulomb problem
\begin{equation}
\widetilde{R}_{n,k}(r)=2\left[\frac{\Gamma(n+1)}{\Gamma\left(n+2k\right)}\right]^{1/2}(2r)^{k-(\mu_1+\mu_2+1/2)} e^{-r}L_{n}^{2k-1}\left(2r\right). \label{st}
\end{equation}
Since $n=n_r$, from equation (\ref{k}) we obtain
\begin{equation}
\widetilde{R}_{n_r,m}(r)=2\left[\frac{\Gamma(n_r+1)}{\Gamma\left(n_r+4m+2\mu_1+2\mu_2+1\right)}\right]^{1/2}(2r)^{2m} e^{-r}L_{n_r}^{4m+2\mu_1+2\mu_2}\left(2r\right).\label{Rfin}
\end{equation}
By using the relation
\begin{equation}\label{scall}
e^{i\theta{A_2}}f(r)=e^{\theta}f(e^{\theta}r),
\end{equation}
where $f(r)$ is an arbitrary spherically symmetric function, the radial eigenfunctions $R(r)=e^{i\theta A_2}\widetilde{R}(r)$ can be written as
\begin{equation}
R(r)=N_n2\left[\frac{\Gamma(n_r+1)}{\Gamma\left(n_r+4m+2\mu_1+2\mu_2+1\right)}\right]^{1/2}(2ar)^{2m} e^{-ar}L_{n_r}^{4m+2\mu_1+2\mu_2}\left(2ar\right),\label{Rfin2}
\end{equation}
where $a=\sqrt{-2\mathcal{E}}$ and $N_n$ is a normalization constant which is computed by using equation (\ref{norm}). Therefore,
\begin{align}\nonumber
R(r)&=\sqrt{\frac{(2a)^{2\mu_1+2\mu_2+2}n_r!}{\Gamma(n_r+4m+2\mu_1+2\mu_2+1)(2n_r+4m+2\mu_1+2\mu_2+1)}}\times\\
&\times e^{-ar}(2ar)L_{n_r}^{4m+2\mu_1+2\mu_2}(2ar).\label{Rn}
\end{align}
The energy spectrum and radial eigenfunction of equations (\ref{En}) and (\ref{Rn}) coincide with those previously obtained in
Section $3$ by using Schr\"odinger factorization method. In this Section the radial eigenfunctions for the Dunkl-Coulomb problem  were obtained by readapting  the Kepler-Coulomb Sturmian basis reported  by
Gerry and Gur. Our results are also consistent with those obtained in reference \cite{GEN4}.

The Schr\"odinger factorization method and the tilting transformation have been successfully applied to study several relativistic and non-relativistic problems, as can be seen in references \cite{NOS2,NOS3}.

\section{$SU(1,1)$ radial coherent states}

The $SU(1,1)$ Perelomov coherent states $\widetilde{R}(r,\xi)$ for the Dunkl-Coulomb problem for the radial functions
can be computed in a closed form. These states are defined as the action of the displacement
operator $D(\xi)$ on the lowest normalized state $|k,0\rangle$ \cite{PERL}
\begin{equation}
|\zeta\rangle=D(\xi)|k,0\rangle=(1-|\xi|^2)^k\sum_{n=0}^\infty\sqrt{\frac{\Gamma(n+2k)}{n!\Gamma(2k)}}\xi^n|k,n\rangle.
\end{equation}
Thus, from the Sturmian functions $\widetilde{R}(r)$ we obtain
\begin{equation}\label{est1}
\widetilde{R}(r,\xi)=2\left[\frac{(1-|\xi|^2)^{2k}}{\Gamma(2k)}\right]^{1/2}(2r)^{k-(\mu_1+\mu_2+1/2)} e^{-r}\sum_{n=0}^\infty\xi^nL_{n}^{2k-1}\left(2r\right).
\end{equation}
From the Laguerre polynomials generating function
\begin{equation}
\sum_{n=0}^\infty L_n^\nu(x)y^n=\frac{e^{-xy/(1-y)}}{(1-y)^{\nu+1}},
\end{equation}
we obtain that the radial coherent states $\widetilde{R}(r,\xi)$ are
\begin{equation}
\widetilde{R}(r,\xi)=2\left[\frac{(1-|\xi|^2)^{2k}}{\Gamma(2k)(1-\xi)^{4k}}\right]^{1/2}(2r)^{k-(\mu_1+\mu_2+1/2)} e^{\frac{r\left(\xi+1\right)}{\left(\xi-1\right)}}.
\end{equation}

The energy spectrum $\mathcal{E}$ of this problem can be expressed in terms of the coherent parameter $\xi$. From equation (\ref{ecde}), it follows
\begin{equation}
\langle \zeta|\widetilde{H}|\zeta\rangle=\sqrt{-2\mathcal{E}}\langle \zeta|A_0|\zeta\rangle=-\alpha.
\end{equation}
Thus, using equation (\ref{cosh}) of Appendix we obtain
\begin{equation}\label{Ener2}
\mathcal{E}=-\frac{\alpha^2}{2\left(2m+\mu_1+\mu_2+1/2\right)^2\cosh^2\left(2|\xi|\right)}.
\end{equation}

The physical coherent states can be computed from the relationship $R(r,\xi)=e^{i\theta A_2}\widetilde{R}(r,\xi)$ (see equation (\ref{scall})),
\begin{equation}
R(r,\xi)=C_n2\left[\frac{(1-|\xi|^2)^{2k}}{\Gamma(2k)(1-\xi)^{4k}}\right]^{1/2}(2ar)^{k-(\mu_1+\mu_2+1/2)} e^{\frac{ar\left(\xi+1\right)}{\left(\xi-1\right)}},
\end{equation}
where again $a=\sqrt{-2\mathcal{E}}$ and $C_n$ is a normalization constant to be determined. Therefore,
the $SU(1,1)$ physical coherent states for the Dunkl-Coulomb in terms of $m,\mu_1,\mu_2$ are
\begin{equation}
R(r,\xi)=2C_n\left[\frac{(1-|\xi|^2)^{4m+2\mu_1+2\mu_2+1}}{\Gamma(4m+2\mu_1+2\mu_2+1)(1-\xi)^{8m+4\mu_1+4\mu_2+2}}\right]^{1/2}(2ar)^{2m} e^{\frac{ar\left(\xi+1\right)}{\left(\xi-1\right)}}.
\end{equation}
The new normalization constant $C_n$ can be computed as follows
\begin{align}\nonumber\label{CNOR1}
1&=\langle\overline{\xi,\mu_1,\mu_2}|A_0-A_1|\overline{\xi,\mu_1,\mu_2}\rangle\\
&=C^2e^{-\theta}\langle \xi,\mu_1,\mu_2|A_0-A_1|\xi,\mu_1,\mu_2\rangle\\\nonumber
&=C^2e^{-\theta}k\left[\cosh(2|\xi|)+\sinh(2|\xi|)\cos\varphi\right].
\end{align}
Thereby, the normalization constant is
\begin{equation}
C_n=\frac{\left(-2E\right)^{1/2}}{\left(2m+\mu_1+\mu_2+1/2\right)^{1/2}\left[\cosh(2|\xi|)+\sinh(2|\xi|)\cos\varphi\right]^{1/2}}.
\end{equation}

\section{Concluding remarks}

The radial part of the Dunkl-Coulomb problem was studied in polar coordinates by using the $su(1,1)$ Lie algebra
and its theory of irreducible representations. We constructed two different sets of Lie algebra generators
for this problem. The first realization was obtained from the Schr\"odinger factorization
method. In the second realization, the generators are energy-independent and  one of the them is proportional to
the radial Schr\"odinger equation.

The radial functions for each realization were obtained in different ways. For the Schr\"odinger factorization they were
found by solving analytically the differential equation. For the second realization, we used the Sturmian basis of the
$su(1,1)$ Lie algebra. Moreover, this Sturmian basis of the Lie algebra was used to calculate the Perelomov coherent states
for the radial part. Finally, we used these coherent states and the $su(1,1)$ theory to find  the energy spectrum and the
normalization constant in terms of the coherent states parameter.

It is important to note that in reference \cite{GEN4}, the authors also used a realization of $so(2,1)$ in terms of Dunkl operators
to derive the spectrum of this problem. However, their generators are very different to those introduced in the presented work.

\section*{Acknowledgments}
This work was partially supported by SNI-M\'exico, COFAA-IPN,
EDI-IPN, EDD-IPN, SIP-IPN project number $20161727$.

\renewcommand{\theequation}{A.\arabic{equation}}
\setcounter{equation}{0}
\section*{Appendix: The $SU(1,1)$ Group and its coherent states}

The $su(1,1)$ Lie algebra is spanned by the generators $K_{+}$, $K_{-}$
and $K_{0}$, which satisfy the commutation relations \cite{VOU}
\begin{eqnarray}
[K_{0},K_{\pm}]=\pm K_{\pm},\quad\quad [K_{-},K_{+}]=2K_{0}.\label{com}
\end{eqnarray}
The action of these operators on the basis $\{|k,n\rangle, n=0,1,2,...\}$ is
\begin{equation}
K_{+}|k,n\rangle=\sqrt{(n+1)(2k+n)}|k,n+1\rangle,\label{k+n}
\end{equation}
\begin{equation}
K_{-}|k,n\rangle=\sqrt{n(2k+n-1)}|k,n-1\rangle,\label{k-n}
\end{equation}
\begin{equation}
K_{0}|k,n\rangle=(k+n)|k,n\rangle,\label{k0n}
\end{equation}
where $|k,0\rangle$ is the lowest normalized state. The Casimir
operator for any irreducible representation satisfies
\begin{equation}
K^{2}=-K_{+}K_{-}+K_{0}(K_{0}-1)=k(k-1).\label{cas}
\end{equation}
The theory of unitary irreducible representations of the $su(1,1)$ Lie algebra has been
studied in several works \cite{ADAMS} and it is based on equations (\ref{k+n})-(\ref{cas}).
Thus, a representation of $su(1,1)$ algebra is determined by the number $k$. For the purpose of the
present work we will restrict to the discrete series only, for which
$k>0$.

The $SU(1,1)$ Perelomov coherent states $|\zeta\rangle$ are
defined as \cite{PERL}
\begin{equation}
|\zeta\rangle=D(\xi)|k,0\rangle,\label{defPCS}
\end{equation}
where $D(\xi)=\exp(\xi K_{+}-\xi^{*}K_{-})$ is the displacement
operator and $\xi$ is a complex number. From the properties
$K^{\dag}_{+}=K_{-}$ and $K^{\dag}_{-}=K_{+}$ it can be shown that
the displacement operator possesses the property
\begin{equation}
D^{\dag}(\xi)=\exp(\xi^{*} K_{-}-\xi K_{+})=D(-\xi),
\end{equation}
and the so called normal form of the displacement operator is given by
\begin{equation}
D(\xi)=\exp(\zeta K_{+})\exp(\eta K_{0})\exp(-\zeta^*
K_{-})\label{normal},
\end{equation}
where $\xi=-\frac{1}{2}\tau e^{-i\varphi}$, $\zeta=-\tanh
(\frac{1}{2}\tau)e^{-i\varphi}$ and $\eta=-2\ln \cosh
|\xi|=\ln(1-|\zeta|^2)$ \cite{GER}. By using this normal form of the displacement
operator and equations (\ref{k+n})-(\ref{k0n}), the Perelomov coherent states are found to
be \cite{PERL}
\begin{equation}
|\zeta\rangle=(1-|\xi|^2)^k\sum_{n=0}^\infty\sqrt{\frac{\Gamma(n+2k)}{n!\Gamma(2k)}}\xi^n|k,n\rangle.\label{PCN}
\end{equation}
Now, by using the Baker-Campbell-Hausdorff identity
\begin{equation}
e^{-A}Be^A=B+\frac{1}{1!}[B,A]+\frac{1}{2!}[[B,A],A]+\frac{1}{3!}[[[B,A],A],A]+...,\label{Baker}
\end{equation}
and equation (\ref{com}), we can find the similarity transformations
$D^{\dag}(\xi)K_{+}D(\xi)$, $D^{\dag}(\xi)K_{-}D(\xi)$ and
$D^{\dag}(\xi)K_{0}D(\xi)$ of the $su(1,1)$ Lie algebra generators.
These results are given by
\begin{align}\label{simiK+}
D^{\dag}(\xi)K_{+}D(\xi) &=\frac{\xi^{*}}{|\xi|}\alpha K_{0}+\beta\left(K_{+}+\frac{\xi^{*}}{\xi}K_{-}\right)+K_{+},\\\label{simiK-}
D^{\dag}(\xi)K_{-}D(\xi) &=\frac{\xi}{|\xi|}\alpha K_{0}+\beta\left(K_{-}+\frac{\xi}{\xi^{*}}K_{+}\right)+K_{-},\\\label{simiK0}
D^{\dag}(\xi)K_{0}D(\xi)&=(2\beta+1) K_{0}+\frac{\alpha\xi}{2|\xi|}K_{+}+\frac{\alpha\xi^*}{2|\xi|}K_{-},
\end{align}
where $\alpha=\sinh(2|\xi|)$ and
$\beta=\frac{1}{2}\left[\cosh(2|\xi|)-1\right]$.

Moreover, the expectation values of the group generators $K_{\pm}, K_0$ in the Perelomov number coherent states
can be easily computed by using the similarity transformations, equations (\ref{simiK+})-(\ref{simiK0}).
Thus,
\begin{align}\label{sinh}
\langle \xi,k|K_{\pm}|\xi,k\rangle &=-e^{\pm{i\varphi}}k\sinh(2|\xi|),\\\label{cosh}
\langle \xi,k|K_{0}|\xi,k\rangle &=k\cosh(2|\xi|).
\end{align}


\begin{thebibliography}{99}
\bibitem{YANG}    L.M. Yang, Phys. Rev. {\bf84} 788 (1951).
\bibitem{DUNKL}   C.F. Dunkl, Trans. Am. Math. Soc. {\bf311} 167 (1989).
\bibitem{XuY}     C.F. Dunkl, Y. Xu, \textit{Orthogonal polynomials of several variables, Encyclopedia of Mathematics and Its
                  Applications}, Vol. 81, Cambridge University Press, Cambridge, 2001.
\bibitem{Chere}   L. Cherednik, Invent. Math. {\bf106} 411 (1991).
\bibitem{Opdam}   E.M. Opdam, Acta Math. {\bf175} 75 (1995).
\bibitem{HIK}     K. Hikami, J. Phys. Soc. Japan {\bf65} 394 (1996).
\bibitem{KAK}     S. Kakei, J. Phys. A: Math. Gen. {\bf29} L619 (1996).
\bibitem{LAP}     L. Lapointe, L. Vinet, Comm. Math. Phys. {\bf178} 425 (1996).
\bibitem{GEN1}    V.X. Genest, M.E.H. Ismail, L. Vinet, A. Zhedanov, J. Phys. A: Math. Theor. {\bf46} 145201 (2013).
\bibitem{GEN2}    V.X. Genest, M.E.H. Ismail, L. Vinet, A. Zhedanov, Commun. Math. Phys. {\bf329} 999 (2014).
\bibitem{GEN3}    V.X. Genest, L. Vinet, A. Zhedanov, J. Phys. Conf. Ser. {\bf512} 012010 (2014).
\bibitem{NOS}     M. Salazar-Ram\'irez, D. Ojeda-Guill\'en, R.D. Mota, V.D. Granados, \textit{$SU(1,1)$ solution for the Dunkl oscillator in two dimensions and its coherent states}, arXiv preprint arXiv:1607.06169.
\bibitem{GEN4}    V.X. Genest, A. Lapointe, L. Vinet, Phys. Lett. A {\bf379} 923 (2015).
\bibitem{DIRLIB}  P.A.M. Dirac, \textit{The Principles of Quantum Mechanics}, Clarendon Press, Oxford, 1935.
\bibitem{SCH1A}   E. Schr\"odinger, Proc. R. Ir. Acad. A {\bf46} 9 (1940).
\bibitem{SCH1B}   E. Schr\"odinger, Proc. R. Ir. Acad. A {\bf46} 183 (1940).
\bibitem{SCH1C}   E. Schr\"odinger, Proc. R. Ir. Acad. A {\bf47} 53 (1941).
\bibitem{INF1}    L. Infeld, Phys. Rev. {\bf59} 737 (1941).
\bibitem{INF2}    L. Infeld, T.E. Hull, Rev. Mod. Phys. {\bf23} 21 (1951).
\bibitem{SCH}     E. Schr\"odinger, Naturwiss. {\bf14} 664 (1926).
\bibitem{PERL}    A.M. Perelomov, \textit{Generalized Coherent States and Their Applications}, Springer-Verlag, Berlin, 1986.
\bibitem{ZHANG}   W.M. Zhang, D.H. Feng, R. Gilmore, Rev. Mod. Phys. {\bf62} 867 (1990).
\bibitem{GAZL}    J.P. Gazeau, \textit{Coherent States in Quantum Physics}, Wiley-VCH, Berlin, 2009.
\bibitem{KLIL}    A.B. Klimov, S.M. Chumakov, \textit{A Group-Theoretical Approach to Quantum Optics}, Wiley-VCH, Weinheim, 2009.
\bibitem{AOB}     A.O. Barut, L. Girardello, Commun. Math. Phys. {\bf21} 41 (1971).
\bibitem{AMP}     A.M. Perelomov, Commun. Math. Phys. {\bf26} 222 (1972).
\bibitem{KWO}     K. Wodkiewicz, J.H. Eberly, J. Opt. Soc. Am. {\bf2} 458 (1985).
\bibitem{CBR}     C. Brif, A. Vourdas, A. Mann, J. Phys. A {\bf29} 5873 (1996).
\bibitem{CBR2}    C. Brif, Int. J. Theor. Phys. {\bf36} 1651 (1997).
\bibitem{DANIEL}  D. Mart\'{\i}nez, R.D. Mota, Ann. Phys. {\bf323} 1024 (2008).
\bibitem{LEB}     N.N. Lebedev, \textit{Special Functions and their Applications, Dover Publications}, New York, 1972.
\bibitem{ADAMS}   B.G. Adams, \textit{Algebraic Approach to Simple Quantum Systems}, Springer, Berlin, 1994.
\bibitem{AOB2}    A.O. Barut, \textit{Dynamical Groups and Generalized Symmetries in Quantum Theory}, University of Carterbury Press, New Zeland, 1972.
\bibitem{KTH}     K.T. Hecht, \textit{Quantum Mechanics}, Springer-Verlag, New York, 2000.
\bibitem{gerry}   C.C. Gerry, J. Kiefer, Phys. Rev. A {\bf37}, 665 (1988).
\bibitem{gur}     Y. Gur, A. Mann, Phys. At. Nucl. {\bf68} 1700 (2005).
\bibitem{NOS2}    M. Salazar-Ram\'irez, D. Ojeda-Guill\'en, R.D. Mota, J. Math. Phys. {\bf57} 021704 (2016).
\bibitem{NOS3}    M. Salazar-Ram\'irez, D. Ojeda-Guill\'en, R.D. Mota, Ann. Phys. {\bf372} 283 (2016).
\bibitem{VOU}     A. Vourdas, Phys. Rev. A {\bf41} 1653 (1990).
\bibitem{GER}     C.C. Gerry, Phys. Rev. A {\bf31} 2721 (1985).

\end{thebibliography}
\end{document}